
\magnification=1200
\baselineskip=20pt
\tolerance=100000
\overfullrule=0pt
\centerline{\bf EFFECTS OF NONGAUGE POTENTIALS ON THE}
\centerline{\bf SPIN-1/2 AHARONOV-BOHM PROBLEM}

\vskip 1cm

\centerline{by}

\vskip 1cm

\centerline{C. R. Hagen}
\centerline{Department of Physics and Astronomy}
\centerline{University of Rochester}
\centerline{Rochester, NY 14627}

\vskip 1cm

\centerline{\bf Abstract}

\bigskip

Some recent work has attempted to show that the singular solutions which
are known to occur in the Dirac description of spin-1/2 Aharonov-Bohm
scattering can be eliminated by the inclusion of strongly repulsive
potentials inside the flux tube.  It is shown here that these calculations
are generally
 unreliable since they necessarily require potentials which lead to the
occurrence of Klein's paradox.  To avoid that difficulty
 the problem is
solved within the framework of the Galilean spin-1/2 wave equation which is
free of that particular complication.
  It is then found that the singular solutions can be
eliminated provided that the nongauge potential is made energy dependent.
The effect of the inclusion of a Coulomb potential is also considered with
the result being that the range of flux parameter for which singular
solutions are allowed is only half as great as in the pure Aharonov-Bohm
limit.
Expressions are also obtained for the binding energies which can occur in
the combined Aharonov-Bohm-Coulomb system.

\vfill\eject

\noindent {\bf I. Introduction}
\medskip

In classical physics it is a trivial fact that the absence of a force
 necessarily implies zero scattering.  On the other hand the
 very remarkable Aharonov-Bohm (AB) effect$^1$ shows that this does not
apply in the realm of quantum mechanics and that potentials (as opposed to
fields themselves) can indeed have observable consequences.  Thus charged
particles are found to be scattered by a thin magnetized filament even
though it is possible, by shielding the flux tube or filament, to establish
that penetration into the region of nonvanishing magnetic field cannot
occur.

For the scattering of a nonrelativistic particle of mass $M$ by the
potential
$$e A_i = \alpha\  \epsilon_{ij} r_j / r^2 \eqno(1)$$
where $r_i$ is the radius vector in two dimensions and $\alpha$ is the
flux parameter one needs to solve the Schr\"odinger equation
$${1 \over 2M} \ \left( {1 \over i}\ \nabla - e {\bf A} \right)^2
 \psi = E \psi \quad . \eqno(2)$$
Upon writing
$$\psi (r, \phi) = \sum^\infty_{- \infty} e^{i m \phi} f_m (r) \eqno(3)$$
Eq. (2) reduces to the Bessel equation
$$\left[ {1 \over r} \ {d \over dr}\ r \ {d \over dr} + k -
(m + \alpha )^2 / r^2 \right] f_m (r) = 0 \quad . \eqno(4)$$
Since (4) has both a regular and irregular solution, it is necessary to
give a boundary condition which allows a unique result to be obtained.  One
could, of course,
 simply require that $f_m (r)$ be finite at $r=0$ and thereby
eliminate {\it ab initio} the irregular solution.  This in fact gives the
well known AB solution.  Since, however, a resolution of this issue by fiat
is totally unsuccessful when spin is included, a more physical approach
would clearly be preferable.  This is accomplished$^2$ by replacing (1)
by$^3$
$$e A_i = \cases{\alpha\ \epsilon_{ij} r_j/r^2 & $\ \ \ \ \ r> R$\cr
\noalign{\vskip 6pt}%
0 & $\ \ \ \ \ r< R$\cr}\eqno(5)$$
and taking the limit $R \rightarrow 0$ after matching boundary conditions
at $r=R$.  Clearly, the vector potential (5) mathematically effects
 the replacement of
an idealized zero thickness filament by one of finite radius $R$ which has
a surface distribution of magnetic field given by
$$e H = - {\alpha \over R}\ \delta (r-R) \quad . $$
Actually, the specific details of the model (5) can be shown to be
irrelevant$^3$ provided only that the flux distribution is independent of
angle and has no delta function contribution at the origin.  It is thus
straightforward to establish$^2$ that the irregular solution is absent and
that the usual AB solution obtains in the $R=0$ limit.

When spin is included, the situation becomes quite different.  Here one is
concerned with the (two-component) Dirac equation
$$E \psi = \left[ M \beta + \beta \gamma \cdot \Pi \right] \psi \eqno(6)$$
where $\Pi_i = - i \partial_i - e A_i$ with $A_i$ as in (1).  A convenient
choice for the matrices in (6) is
$$\eqalign{&\beta  = \sigma_3\cr
&\beta \sigma_i = (\sigma_1, s \sigma_2 )\cr}$$
where the $\sigma$'s are the usual Pauli matrices and $s = \pm 1$ for spin
``up" and spin ``down" respectively.  Upon reduction to a second order form
one obtains from (6) the result
$$\left( E^2 - M^2 \right) \psi = \left[
\Pi^2 + \alpha s \sigma_3 \ {1 \over r}\ \delta (r)
\right] \psi$$
which, upon using a partial wave decomposition of the form (3), becomes
$$\left[ {1 \over r}\ {d \over dr}\ r\ {d \over dr}\  +
k^2  -(m + \alpha)^2 /r^2
 - \alpha s \sigma_3 \ {1 \over r}\ \delta (r) \right] f_m = 0$$
where
$$k^2 = E^2 - m^2 \quad .$$
It is significant that the idealization to a zero radius flux tube has
given rise to the complication that the delta function occurs at a singular
point of the differential equation.

Different approaches have been attempted in dealing with this difficulty.
Alford {\it et al}.$^4$
 simply required the upper component to be regular at the
origin.  This essentially makes the spin term trivial and implies an
amplitude which coincides with the spinless AB result.  This contradicts
the helicity conservation which is known to be valid$^5$ for the system
described by (6).  On the other hand, Gerbert$^6$ has taken an approach
which essentially states that an arbitrary linear combination of the two
solutions $J_{|m+\alpha|} (kr) \ {\rm and}\ J_{-|m+ \alpha|} (kr)$
 in the region $r > R$ can be
taken so long as it is normalizable
 for $R \rightarrow 0$.  Thus when $|m + \alpha| < 1$ an
arbitrary parameter $\theta$ appears which describes the relative
contribution of these two functions.  This somewhat mathematical approach
consequently suffers from the appearance of a parameter with no clear physical
origin.

In ref. 3 an approach was taken which was based on the physically
reasonable modification (5) of the vector potential.  Upon matching
boundary conditions at $r=R$ and letting $R \rightarrow 0$ it was found
that the correct solution was always the regular one except when $\alpha s
< 0$ and $|m + \alpha |< 1$.   In that case there would always be one and
only one allowed irregular solution.  This occurred for
$$m = -N \quad , \quad N \geq 0 \eqno(7)$$
when $s = -1$ and
$$m = -N -1 \quad , \quad N + 1 \leq 0 \eqno(8)$$
for $s = +1$ with the integer $N$ defined by
$$\alpha = N + \beta$$
where
$$0 \leq \beta < 1 \quad .$$
It is interesting to note that in the cases (7) and (8) for which irregular
solutions are allowed there is \underbar{no} contribution from the regular
solution $J_{|m + \alpha |} (kr)$.  Thus the solution obtained in ref. 3
does correspond to a solution of the type obtained by Gerbert without,
however, the introduction of his mixing parameter.  Consequently the
solution is unique and can be shown$^5$ to be consistent with helicity
conservation.  A point worth mentioning is that this approach was motivated
by a desire to formulate the problem from the outset in a physically
meaningful way and \underbar{not} (as recently stated$^7$) to provide (after
the fact) a physical motivation by which to determine Gerbert's
$\theta$ parameter.

It is clear simply from the $\alpha s < 0$ condition that the solution of
ref. 3 is not anyonic.  This has led to some serious concerns by those
who consider the anyonic properties of the spinless AB system to be of
fundamental significance.  Thus, for example, ref. 7 has taken the
introduction of a boundary at $r=R$ one step further by including a
nonvanishing nongauge potential for $r<R$ and examining its effects when it
diverges in certain ways.  In the following section this work is
examined and it is shown that Klein's paradox makes that approach
unreliable.  If, on the other hand, a Galilean spin-1/2 wave equation is used
in place of the Dirac equation which it closely resembles, then the
appearance of Klein's paradox can be avoided.  This leads to the
determination of an ``inside" potential which allows one to force both
regular and irregular solutions to occur.  However, it implies an energy
dependent potential, a fact which is easily seen to follow from dimensional
considerations.  Section III considers the case in which a $\xi /r$
potential is also included.  Remarkably, it is found that for arbitrarily
small $\xi$ the domain of $\alpha$ for which singular solutions can be
obtained now shrinks to $|m + \alpha | < 1/2$.  This contrasts sharply with
the claims of ref. 7, illustrating again the pitfalls of dealing with
equations in which Klein's paradox is known to occur.

\medskip

\noindent {\bf II. Klein's Paradox and How to Avoid It}

\medskip

It was proposed in ref. 7 that one modify the spin-1/2 AB problem as done
in ref. 3 by including for $r < R$ a constant potential $u_R$.  Since
there is no flux for $r < R$, one has to consider in that domain the
equation
$$\left\{ {1 \over r}\ {d \over dr}\ r\ {d \over dr} + \left[
\left( E - u_R \right)^2 - M^2 \right] -
m^2 /r^2 \right\} f_m (r) = 0$$
which has the allowed solution $J_m (k_0 r)$ where
$$k^2_0 = \left( E- u_R \right)^2 - M^2 \quad .$$
Since all the results of ref. 7 depend upon the limit of $u_R \rightarrow
 \infty$, it is of interest to note the behavior of $k_0$ as a function of
$u_R$.  As $u_R$ increases from zero $k_0$ is a real number $(E > m)$
which eventually vanishes at $u_R = E - m$.  At this point $k_0$ becomes
imaginary and the function $J_m (k_0r)$ goes from sinusoidal oscillation to
a real exponential (i.e., $\exp [|k_0|r]$).  As $u_R$ continues to
increase $k_0$ vanishes again at $u_R = E + m$ and the function $J_m(k_0
r)$ is again oscillatory and remains so as $u_R$ increases without limit.
All of these features are reminiscent of the phenomenon of Klein's
paradox$^8$ and one concludes that the Dirac equation is not an adequate
framework for this problem in the limit of arbitrarily large $u_R$.

On the other hand the type of questions raised in ref. 7 \underbar{are}
amenable to treatment provided that one substitutes for the Dirac equation
its Galilean limit.  This is  given by
$$\left[ {\cal E} {1 \over 2} \ (1 + \beta) + M (1 - \beta)
- \beta \gamma \cdot \Pi \right] \psi = 0 $$
where ${\cal E} = E- M$ is the ``nonrelativistic" energy.  The above was
 derived in ref. 2 from the Dirac equation and is simply the $2 + 1$ space
version of results obtained by L\'evy-Leblond$^9$.  It is easily seen to
imply for $\psi_1$ (the upper component of $\psi$) the result
$$\left[ {\cal E} - {1 \over 2M}\ \Pi^2 - {\alpha s \over 2M} \
{1 \over R}\ \delta (r-R) \right]  \psi_1 = 0 $$
provided that one takes a vector potential of the form (5).  Again,
carrying out a partial wave expansion, one obtains the radial equation
$$\left\{ {1 \over r}\ {d \over dr}\ r\ {d \over dr} + k^2
- \left[ m + \alpha \theta (r-R) \right]^2 /r^2 -
\alpha s {1 \over R} \ \delta (r-R)
\right\} f_m (r) = 0 \eqno(9)$$
where $k^2 = 2 M {\cal E}$ and the step function $\theta (x) \equiv {1
\over 2} \left[ 1 + {x \over |x|} \right]$ has ben introduced for
conciseness.

One can now include the effect of a repulsive potential as considered in
ref. 7.  Upon letting ${\cal E} \rightarrow {\cal E} - u_R$ one sees
that the solution for $r < R$ is $J_m (k_0 r)$ with
$$k^2_0 = 2 M ({\cal E} - u_R) \quad .$$
At $u_R = 0$ this implies an oscillatory $J_m (k_0 r)$ which becomes a real
exponential for all $u_R > {\cal E}$.  In other words the wave function
is exponentially damped  in the expected quantum
mechanical fashion
 as $r$ decreases from $R$.  The crucial point is that there is no subsequent
transition back to oscillatory
behavior as $u_R$ increases without limit and
consequently Klein's paradox has been
eliminated.  Thus Eq. (9) is seen to be an appropriate vehicle for carrying
out the program of ref. 7 which seeks to determine whether a suitably
repulsive $u_R$ can allow the simultaneous occurrence of both
$J_{|m + \alpha |} (kr)$ \underbar{and} $J_{-|m+ \alpha |} (kr)$ for
$r > R$.

To carry out this study one writes
$$\eqalign{f_m (r < R) &= I_m (k_0 r)\cr
f_m (r > R) &= A J_{|m + \alpha |} (kr) + BJ_{-|m + \alpha|} (kr)\cr}$$
where $I_m$ is the usual Bessel function of imaginary argument and it has
been assumed that $u_R > {\cal E}$.  The boundary conditions
$$\eqalign{&I_m (|k_0| R) = A J_{|m + \alpha|} (kR) + BJ_{-|m + \alpha|} (kR)
\cr
&R\ {\partial \over \partial r}\ \bigg[ AJ_{|m+ \alpha|} (kr) +
BJ_{-|m + \alpha |} (kr) - I_m (|k_0| r) \bigg]_{r = R} =
\alpha s I_m (|k_0| R)\cr}$$
can be solved to yield for $A/B$ the result
$${A \over B} = {-|m| + g(|k_0|R) \over
2|m + \alpha | + |m| - g (|k_0| R)} \ (kR)^{-2 |m + \alpha|} \eqno(10)$$
where
$$g(x) = x {\partial \over \partial x} \log I_m (x)$$
and use has been made of the relation
$$|m| + |m + \alpha | = - \alpha s$$
since as shown in ref. 3 only in this partial wave is a singular solution
possible.

One can now answer the question posed in ref. 7 whether a finite nonzero
$A/B$ is possible.  It is, however, clear that since $u_R \rightarrow
\infty$ must yield $B/A = 0$ just as $u_R = 0$ gives $A/B = 0$, there
 \underbar{must} exist a value for $u_R$ which implies a finite $A/B$.  The
real issue it would seem is whether a $u_R$ can be found which is
independent of the energy ${\cal E}$.  It is not difficult to show (most
trivially, by dimensional considerations) that no such energy independent
solutions exist.  This is in marked contrast to the results claimed in ref.
7.  In that work the scale for the potential $u_R$ is determined by the
mass $M$, or, in other words, by a factor which unavoidably requires special
relativity.  Since Klein's paradox has been seen to make that approach
unreliable, the scale for $u_R$ is determined necessarily by the
nonrelativistic energy ${\cal E}$.

The (energy dependent) solution of (10) given by
$$|k_0| R = \lambda (kR)^{| m + \alpha |} \eqno(11)$$
where $\lambda$ is arbitrary implies for $R \rightarrow 0$ that
$$A/B = {1 \over 4}\ {\lambda \over | m + \alpha |} \ {1 \over
|m| + 1} \quad . $$
Thus any (positive) value of this ratio can be obtained by appropriate
choice of $\lambda$.  Somewhat curiously, negative values of $A/B$ can also
be generated provided that $u_R$ is attractive but diverges according to
the same power law as in (11).  Thus an affirmative answer has been found
for the issue of fine tuning raised in ref. 7.  The fact that this tuning
requires an intricate dependence of the interior potential on the energy
seems to imply, however, that it can have little, if any, utility.
\vfill\eject
\medskip

\noindent {\bf III. Coulomb Modifications}

\medskip

It is known that partial wave solutions for the AB problem can be obtained
exactly even when a $1/r$ potential is included.  In particular such
solutions have been obtained by Law
{\it et al}.$^{10}$ for the spinless case.
This generalization has also been included in ref. 7 in the context of
their interior repulsive potential $u_R$.  Since it has been remarked
already that Klein's paradox adversely affects such calculations, it is of
interest to describe the results obtained when the Galilean spin-1/2
equation is employed for such an analysis.

Upon taking the potential to be
$$V (r) = \cases{u_R & $\ \ \ \ \ r<R$\cr
\noalign{\vskip 6pt}%
\xi/r & $\ \ \ \ \ r>R$\cr}\quad ,$$
one finds that the appropriate wave equations for the individual partial
waves in the expansion of the upper component of $\psi$ are
$$\left[ {1 \over r}\ {d \over dr}\ r\ {d \over dr} + k^2_0 -
 {m^2 \over r^2} \right] f_m (r) = 0$$
for $r < R$, and
$$\left[ {1 \over r}\ {d \over dr}\ r\ {d \over dr} + k^2 -
 2 M \xi / r - {(m + \alpha)^2  \over r^2} \right] f_m (r) = 0$$
for $r > R$.  It is straightforward to obtain the solution
$$f_m (r) = J_m (k_0 r)$$
for $r < R$, while for $r > R$
$$\eqalign{f_m (r) = &A_m e^{ikr} (-2 ikr)^{|m + \alpha |}
F \bigg( |m + \alpha | + 1/2 +
i M \xi / k \bigg| 2 | m + \alpha | + 1 \bigg|
- 2 ikr \bigg)\cr
\noalign{\vskip 4pt}%
&+ B_m e^{ikr} (-2ikr)^{- |m + \alpha|}
F \bigg( -|m + \alpha | + 1/2 + i M \xi /k
\bigg| 1 - 2 | m + \alpha | \bigg| - 2 ikr \bigg) \cr}$$
where $F(a|c|z)$ is the
usual confluent hypergeometric function.  Note that
$A_m$ and $B_m$ are the coefficients of the regular and irregular solutions
respectively.  It is worth remarking that because of possible modifications
which could be induced by the Coulomb term no assumptions have been made as
to which partial waves can have irregular solutions.

Upon applying the boundary conditions at $r=R$ there obtains
$$\eqalign{{A_m \over B_m} = \bigg\{ &J_{|m|} (k_0 R) R\ {\partial \over
\partial R}\ e^{ikR} (-2ikR)^{-|m + \alpha|}
F \bigg( - |m + \alpha | + 1/2 + i M \xi / k \bigg|
1 - 2 |m + \alpha | \bigg|\cr
\noalign{\vskip 4pt}%
&-2ikr \bigg) -  e^{ikR} (-2ikR)^{-|m + \alpha|}
F \bigg( - |m + \alpha | + 1/2 + i M \xi / k \bigg|
1 - 2 |m + \alpha | \bigg| - 2 ikR \bigg)\cr
\noalign{\vskip 4pt}%
&\bigg( \alpha s + R \ {\partial \over
\partial R} \bigg) J_{|m|} (k_0 R) \bigg\}
\bigg\{  e^{ikR} (-2ikR)^{|m + \alpha|}
F \bigg(  |m + \alpha | + 1/2 + i M \xi / k \bigg|\cr
\noalign{\vskip 4pt}%
& 2 | m + \alpha | + 1 \bigg| -2ikR \bigg)
\bigg( \alpha s + R \ {\partial \over \partial R}
\bigg) J_{|m|} (k_0 R) - J_{|m|} (k_0 R)
R\ {\partial \over \partial R}\cr
\noalign{\vskip 4pt}%
&e^{ikR} (-2 i kR)^{|m + \alpha |} F \bigg( |m + \alpha | + 1/2
+ i M \xi /k \bigg| 2|m + \alpha | +1 \bigg|
-2i kR \bigg) \bigg\}^{-1} \cr}\quad . \eqno(12)$$
One now takes the $R \rightarrow 0$ limit and finds (as in the $\xi =
 0$ case) that $B_m$ must vanish unless $\alpha s < 0$, and
$$|m| + |m + \alpha| = - \alpha s \quad . \eqno(13)$$
However, Eq. (13) is necessary but has not been shown to be
 sufficient.  Nor can one
merely assume on the basis of the analysis of refs. 3 and 6 that $|m
 + \alpha | < 1$.  As stressed in the former work a second condition emerges
when one considers the next-to-leading term in powers of $R$.  In the $\xi
= 0$ case things are considerably simpler since the solutions for both $r <
R$ and $r > R$ are Bessel functions whose expansions are characterized by
the fact that only alternate powers of the argument occur.  This is
 \underbar{not} true for the confluent hypergeometric function and one
finds from (12) that the expansion in powers of $R$ yields
$${A_m \over B_m} \sim \xi R^{1-2 |m + \alpha |} \eqno(14)$$
for small $R$.  Thus singular solutions are possible for $\xi \not= 0$ only
when $| m + \alpha | < 1/2$ rather than the full range $|m + \alpha | < 1$
assumed in ref. 7.  For the case $\xi = 0$ the result (14) is replaced by
$${A_m \over B_m} \sim R^{2-2 |m + \alpha |} \eqno(15)$$
because of the noted property of the Bessel function for small argument.
Clearly, the condition $|m + \alpha | < 1$ follows from (15) whereas the
considerably stronger condition $|m + \alpha | < 1/2$ is required when a
Coulomb term is present.
It should be remarked that when this condition is satisfied, the result
(11) again obtains to determine the dependence of $u_R$ on $k$ and $R$ in
the case that both regular and irregular solutions are required to occur.

The energies ${\cal E}_n$ of the bound states which occur for $\xi < 0$ are
readily determined from the series expansions of the relevant confluent
hypergeometric functions.  These yield for $u_R = 0$ the results
$${\cal E}_n = - {1 \over 2}\ {M \xi^2 \over [n - {1 \over 2}\ \pm
|m + \alpha |]^2} \quad , \quad n = 1,2, \dots$$
where the upper and lower signs refer respectively to the case of regular
and irregular solutions.  Of particular interest is the fact that the
binding energies of the irregular solutions become arbitrarily large as $|m
+ \alpha |$ approaches ${1 \over 2}$.  This illustrates the crucial role
played by the condition $|m + \alpha | < {1 \over 2}$ and lends added
credence to the derivation presented above.

\medskip

\noindent {\bf IV. Conclusion}

\medskip

This paper has explored the possibility that more  general solutions to the
spin-1/2 AB problem can be found.  As in the earlier work of ref. 7 this
has been done by introducing a very short range, but singular, repulsive
force.  Not surprisingly, it has been found (at least in the unambiguous
Galilean case)  that such solutions can in fact
 exist.  This is physically
reasonable since the effect of such a potential is to reduce significantly the
interaction of the magnetic moment with the singular magnetic field at the
origin.  On the other hand such a potential must be required to be energy
dependent, and it also has the disadvantage of violating helicity
conservation, a property which would otherwise be satisfied.  Since the
solution$^3$ of this problem without the additional non-gauge potential is
known to be at variance with anyon features, it is clear that those
properties can be restored only with considerable difficulty.  If one
elects to do this, the effect is to negate the full dynamical participation of
the spin in the interaction.

One of the most interesting results obtained here has to do with the
modifications associated with the inclusion of a Coulomb term.  It was
found that the condition $|m + \alpha | < 1$ which is generally thought to
follow from a condition of normalizability of the solution is not
sufficient.  A more careful analysis shows that only half that range is in
fact allowed for singular solutions.  This has as an immediate consequence
that if one considers a gas of such particles then the discontinuities
known to characterize the second virial coefficient$^{11}$
 $B_2 (\alpha , T)$ are
shifted from integer values of $\alpha$ to half-integer values.
This is particularly noteworthy because of the fact that the transition
point has no dependence on the strength of the Coulomb potential.  Thus one
can imagine this parameter to be continuously decreased to zero and find
that the discontinuities in $B_2 (\alpha , T)$
generally occur at half-integers but
at integral values when the Coulomb term exactly vanishes.  This provides a
most remarkable example of a system in which a point of discontinuity of a
variable which has a macroscopic discontinuity has itself a discontinuous
dependence on a microscopic parameter.  This is a subject which clearly
merits additional study.
\medskip
\noindent {\bf Acknowledgment}

\medskip

This work was supported in part by Department of Energy Grant no.
DE-FG-02-91ER40685.

\vfill\eject

\noindent {\bf References}

\medskip

\item{1.} Y. Aharonov and D. Bohm, Phys. Rev. {\bf 115}, 485 (1959).

\item{2.} C. R. Hagen, Int. J. Mod. Phys. {\bf A6}, 3119 (1991).

\item{3.} C. R. Hagen, Phys. Rev. Lett. {\bf 64}, 503 (1990).

\item{4.} M. G. Alford and F. Wilczek, Phys. Rev. Lett. {\bf 62}, 1071
(1989).

\item{5.} C. R. Hagen, Phys. Rev. Lett. {\bf 64}, 2347 (1990).

\item{6.} Ph. Gerbert, Phys. Rev. D {\bf 40}, 1346 (1989).

\item{7.} F. A. B. Coutinho and J. F. Perez, Phys. Rev. D {\bf 48}, 932
(1993).

\item{8.} A detailed but elementary description of Klein's paradox is given
on pages 40-42 of {\bf Relativistic Quantum Mechanics} by J. D. Bjorken and
S. D. Drell, Mc-Graw Hill (1964).

\item{9.} J. M. L\'evy-Leblond, Commun. Math. Phys. {\bf 6}, 286 (1967).

\item{10.} J. Law, M. K. Srivastava, R. K. Bhaduri, and A. Khare, J. Phys.
A {\bf 25}, L183 (1992).

\item{11.} T. Blum, C. R. Hagen and S. Ramaswamy, Phys. Rev. Lett.
{\bf 64}, 709 (1990).  It is rather odd that this paper (which deals not at
all with spin one) should be cited in ref. 7 as support for its claims that the
spin one and spin-1/2 results are essentially identical.  In fact as shown
by C. R. Hagen and S. Ramaswamy, Phys. Rev. D {\bf 42}, 3524 (1990) the
relativistic spin one problem has no satisfactory AB solution of its wave
equation.

\end